# Magnetic Field-Controlled Mixed Modulation in Magnetoelectric Sensors

*Elizaveta Spetzler\*, Benjamin Spetzler, Dennis Seidler, Johan Arbustini, Lars Thormählen, Robert Rieger, Andreas Bahr, Dirk Meyners, Jeffrey McCord*

E. Spetzler, D. Seidler, J. McCord

Nanoscale Magnetic Materials - Magnetic Domains, Department of Materials Science, Kiel University, Kiel, Germany

E-mail: elgo@tf.uni-kiel.de

B. Spetzler

Energy Materials and Devices, Department of Materials Science, Kiel University, Kiel 24143, Germany

Micro- and Nanoelectronic Systems, Institute of Micro and Nanotechnologies MacroNano, Technische Universität Ilmenau, Ilmenau 98693, Germany

J. Arbustini, R. Rieger

Networked Electronic Systems, Department of Electrical Engineering and Information Technology, Kiel University, Kiel 24143, Germany.

L. Thormählen, D. Meyners

Inorganic Functional Materials, Department of Materials Science, Kiel University, Kiel 24143, Germany

D. Meyners, J. McCord

Kiel Nano, Surface and Interface Science (KiNSIS), Kiel University, Christian-Albrechts-Platz 4, 24118 Kiel, Germany

A. Bahr

Biomedical Electronics, Technische Universität Dresden, 01069 Dresden, Germany.

Magnetoelectric (ME) magnetic field sensors commonly rely on one of the two modulation principles: the nonlinear dependence of magnetostrictive strain on the applied field or the stress-induced change in magnetization susceptibility. While both effects coexist in any ME device, different readout schemes can be chosen to utilize one or the other effect for magnetic field sensing. This work demonstrates that both principles can be simultaneously implemented in a single electrically modulated ME sensor with inductive readout (a converse ME sensor). This mixed modulation approach significantly enhances low-frequency sensitivity while not affecting the sensitivity at higher frequencies. This leads to a nontrivial dependency of the sensor sensitivity on the frequency of the magnetic field to be measured and can effectively decrease the sensor bandwidth by up to an order of magnitude. We show that the contribution of the modulation from the nonlinearity of the magnetostrictive strain to the sensor sensitivity can be changed by applying a magnetic bias field, offering an additional dimension to the design of ME sensors, especially for potential applications in the unshielded environment.

## 1. Introduction

Magnetic field sensors play a crucial role in a wide range of modern applications, including wireless communication systems[1] and biomedical diagnostics.[2] The latter often require detection of extremely weak magnetic fields at low frequencies below 10 kHz with a bandwidth of up to 1 kHz. [3] Recent decades saw substantial progress in the development of magnetic field sensors that target these applications, including sensor concepts utilizing magnetoelectric (ME) composite materials.[4–6] Such materials comprise mechanically coupled piezoelectric and magnetostrictive components, allowing the conversion of magnetic fields into electrical voltage (direct ME effect) and vice versa (converse ME effect) with a much higher conversion factor in comparison to the single-phase ME materials.[7]

Various magnetic field sensor concepts based on the ME effect have been presented; many of them utilize resonator structures, such as cantilevers.[4] A typical ME cantilever consists of three main components: a piezoelectric layer sandwiched between two electrodes, a silicon substrate, and a magnetostrictive layer on the rear side of the substrate. The sensitivity of the sensors based on the direct ME effect is significantly amplified when the frequency $f_{ac}$ of the magnetic field to be measured $H_{ac}$ matches the resonance frequency $f_r$ of the sensor element. Extremely low limits of detection of a few pT/Hz½ have been achieved in this operation mode,[8] however, restricted to narrow-band signals near $f_r$.

Modulation techniques based on the nonlinearity of magnetostriction have been employed to detect low-frequency, low-amplitude magnetic fields over broader frequency ranges.[9] Examples include concepts based on the ΔE and converse ME effects.[4] These concepts correspondingly employ the dependency of the magnetostrictive strain[10] and the susceptibility of magnetization to stress[11] on the applied magnetic field. Although both modulation techniques have been comprehensively studied in the literature[9–11,12,13], the case of mixed modulation and its effect on the signal and bandwidth of converse ME sensors was not investigated. Yet, it is expected that the ΔE effect in the magnetostrictive layer will lead to the mixed modulation of magnetization oscillation in converse ME sensors and, by that, make some contribution to the sensitivity of such sensors. Here, we explain how significant this contribution is and how it changes depending on the applied magnetic bias field. Furthermore, we demonstrate its crucial effect on the sensor sensitivity as a function of the frequency of the measured field, which ultimately determines the bandwidth of the magnetic field sensor and results in a complex dependency of the bandwidth on the bias field.

## 2. Results

### 2.1. Sensor Concept and Mixed Modulation Principle

*2.1.1. Modulation Principle based on the ΔE effect*

The ΔE effect manifests in the dependency of the stiffness tensor of the magnetostrictive material on the magnetostrictive strain and its nonlinear dependency on the magnetic field.[10] In the case of cantilever-based sensors, the change in the stiffness of the magnetostrictive layer causes a change of $f_r$, which can be read out optically,[14] electrically,[15,16] or inductively.[17]

For low amplitudes of an alternating field $H_{ac}$, oscillation of $f_r$ around $f_r(H_{ac} = 0)$ can be approximated as

$$f_r(t) = f_r(H_{ac} = 0) + \frac{\partial f_r}{\partial H} H_{ac}(t), \tag{1}$$

with the time $t$. The shift of $f_r$ with $H_{ac}$ can be calculated via the change $\Delta S$ of the compliance tensor $S$:[13]

$$S_{ij} = \frac{\partial(\varepsilon_i + \lambda_i)}{\partial \sigma_j} = S_{0,ij} + \Delta S_{ij}, \quad i,j \in \{1 \ldots 6\}, \tag{2}$$

$$\frac{\partial f_r}{\partial H} \propto \frac{\partial S_{ij}}{\partial H} = \frac{\partial}{\partial H} \frac{\partial \lambda_i}{\partial \sigma_j}$$

where $\varepsilon$ is the mechanical strain. For the bending modes, $f_r$ is determined by $S_{ii}$. Hence,

$$\frac{\partial f_r}{\partial H} \propto \frac{\partial S_{ii}}{\partial H} = \frac{\partial}{\partial H} \frac{\partial \lambda_i}{\partial \sigma_i} = 3\lambda_s \frac{\partial m_i \chi_{ii}^\sigma}{\partial H}, \tag{3}$$

where $\chi_{ii}^\sigma := \partial m_i / \partial \sigma_i$ is the susceptibility of magnetization to stress and $\lambda_s$ is saturation magnetostriction constant. The sensitivity of the magnetic field sensors based on the $\Delta E$ effect is determined by the change of its resonance frequency, which means that the mechanical bandwidth of the cantilever determines the sensor bandwidth.[15]

*2.1.2. Modulation Principle based on the Converse ME Effect*

An alternative approach that allows broadband signal detection is based on the converse ME effect and the inductive readout of the magnetization change. An illustration of a converse ME sensor and the coordinate system used in this letter is shown in the inset of **Figure 1**a. In this operation mode, the cantilever is excited at its mechanical resonance by applying the excitation voltage $u_{ex}$ between the electrodes with a frequency $f_{ex} = f_r$. The induced stress $\boldsymbol{\sigma}(t) \propto u_{ex}(t)$ in the magnetostrictive layer causes the oscillation of magnetization $m$. If the amplitude of $\boldsymbol{\sigma}(t)$ is small, the oscillation of $m$ can be described via the susceptibility $\chi_{ij}^\sigma := \partial m_i / \partial \sigma_j$. The external magnetic field $H_{ac}(t)$ modulates $\chi_{ij}^\sigma$ leading to the modulation of $m$ as illustrated in Figure 1b. The resulting oscillation of magnetization can be expressed as[11]

$$m_i(t) = \left[\chi_{ij}^\sigma + \frac{\partial \chi_{ij}^\sigma}{\partial H_l} H_{ac,l}(t)\right] \sigma_j(t). \tag{4}$$

This oscillation of magnetization is then read out inductively with a pickup coil.[18,19] This sensor concept significantly differs from the other ME sensor concepts because in this case, the readout signal is determined by the magnetization change and not the magnetostrictive strain. This removes the restriction on the sensor bandwidth imposed by resonator bandwidth. Reported values of the sensor bandwidth can reach several kilohertz, which is ten times larger than the corresponding resonator bandwidth.[6,11] In the following, we show that this is true only within a narrow range of the operation parameters, within which the ΔE effect is negligible.

Because of the geometry of the sensor element and the used resonance mode, often referred to as the U-mode,[5,11,19,20] we can consider only the $x_1$-axis components of $m$ and $H$, and the $x_2$-axis component of $\sigma$ and $\lambda$, i.e., $i = 1, j = 2, l = 1$ in Eq. (4) and $i = 2$ in Eq. (3).[11]

Oscillation of magnetization is recorded using a pickup coil connected to a low-noise operational amplifier, which is then connected to the input of a lock-in amplifier. The same lock-in amplifier provides an auxiliary output for sensor excitation. Details about the setup have been presented elsewhere.[5,11,19] The resulting output voltage $u_{\text{out}}(t)$ at the input of the lock-in amplifier can be approximated as

$$u_{\text{out}}(t) \approx A \cdot \partial m(t)/\partial t, \tag{5}$$

where $A$ is a proportionality factor determined by the geometry of the resonator, the pickup coil, and readout electronics. For the device investigated here, $A \approx 15.4$ mVs. The detailed derivation and validation of Eq. (4-5) and the coefficient $A$ is provided in Ref. [11] as a part of the comprehensive sensor system model.

The first term in Eq. (4) describes the oscillation of magnetization caused by the applied stress $\sigma$ and determines the root-mean-square (RMS) amplitude $U_c$ of the carrier signal $u_c = \sqrt{2}\, U_c \sin(2\pi f_{\text{ex}} t)$ (denoted on the right in Figure 1b). The second term in Eq. (4) describes the modulation of $\chi_{ij}^\sigma$ caused by $H_{\text{ac}}$. It represents the modulation of the carrier and the corresponding amplitude $U_s$ of the sidebands visible in the frequency domain, as illustrated in Figure 1b. Because $U_s$ is linearly proportional to $\partial \chi_{12}^\sigma/\partial H_1$, the dependency of $U_s$ on the bias field $\mu_0 H_{\text{dc}}$ can be qualitatively analyzed by calculating the dependency of $\partial \chi_{12}^\sigma/\partial H_1$ on the bias field $\mu_0 H_{\text{dc}}$. The dependency calculated using a macrospin model is plotted on the left vertical axis of Figure 1a. Details of the model and material parameters used are provided in the Methods.

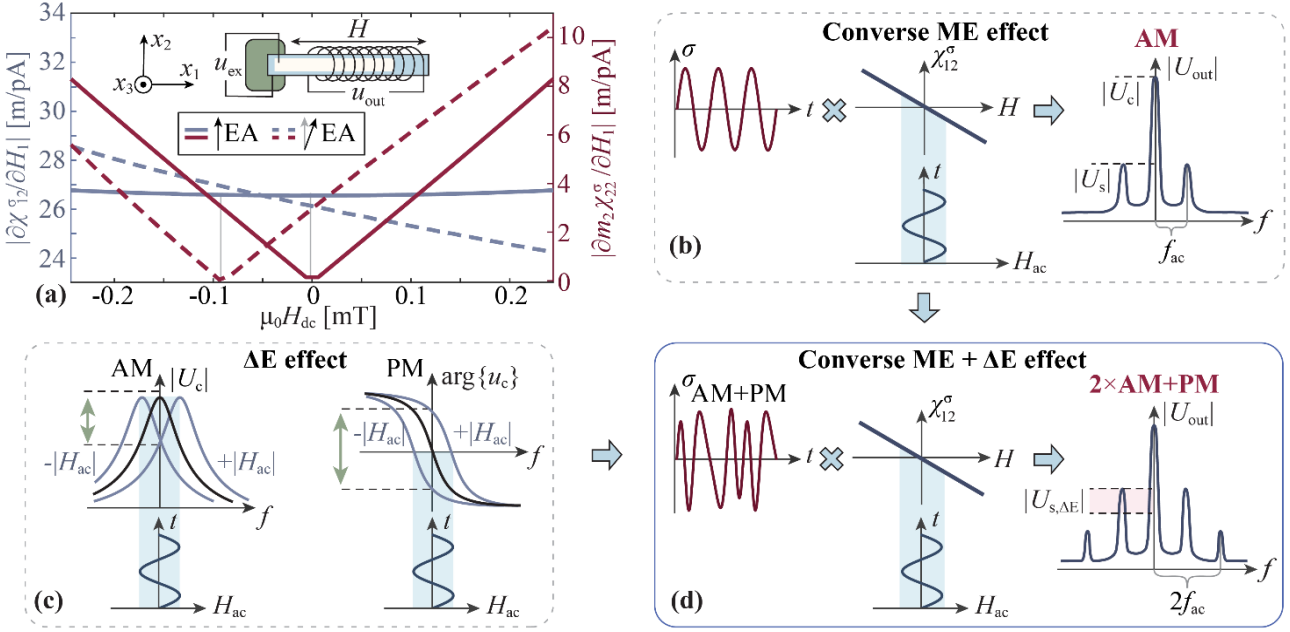

Figure 1. Illustration of the sensor concept. (a) Calculated dependency of $\partial \chi_{12}^\sigma/\partial H_1$ and $\partial m_2 \chi_{22}^\sigma/\partial H_1$ that correspondingly determine the sensitivity of converse ME and ΔE-effect sensors on the magnetic bias field $\mu_0 H_{dc}$. The dependencies are calculated for two orientations of the easy axis (EA) of the uniaxial magnetic anisotropy: along the $x_2$-axis (solid lines) and tilted by 2.6° relative to the $x_2$-axis (dashed lines). Inset: the coordinate system used throughout the paper and a schematic illustration of the sensor element comprising the ME cantilever and the pickup coil. (b) Converse ME effect: applied stress $\sigma$ induces oscillation of magnetization proportional to $\chi_{12}^\sigma$ modulated by $H_{ac}$. The amplitude spectrum of the signal $u_{out}$ read out with the pickup coil is shown on the right. The main peak in the center is the carrier with an amplitude $|U_c|$, the amplitude of the sidebands $|U_s|$ determines the sensitivity of the sensor to the magnetic field. This case illustrates amplitude modulation (AM) of the carrier. (c) ΔE effect: $H_{ac}$ shifts the frequency dependency of the carrier magnitude $|U_c|$ and phase $\arg\{u_c\}$. This leads a amplitude (AM) and phase (PM) modulation of $u_c$ illustrated with the green arrows. (d) Mixed modulation case. The ΔE effect causes amplitude and phase modulation (AM+PM) of $\sigma$, which induces oscillation of $M \propto \chi_{12}^\sigma$ modulated by $H_{ac}$. Amplitude spectrum of the output signal is shown on the right. AM of $\sigma$ creates harmonic distortion, i.e., additional signal peaks at $f_{ex} \pm 2f_{ac}$, PM of $\sigma$ increases the main signal peak $|U_s|$ by adding additional contribution $|U_{s,\Delta E}|$.

The easy axis (EA) of uniaxial magnetic anisotropy is set along the $x_2$-axis (solid lines) and tilted by 2.6° relative to the $x_2$-axis (dashed lines). The tilt of the EA is introduced to mimic the limited manufacturing control and a possible slight deviation from the exact orientation of EA along the $x_2$-axis. Figure 1a shows that if the EA is parallel to the $x_2$-axis, $\partial \chi_{12}^\sigma/\partial H_1$ and, hence, the sensitivity of the converse ME sensor should stay constant within a certain range of $\mu_0 H_{dc}$. The tilt of EA causes the tilt of the $\partial \chi_{12}^\sigma/\partial H_1$ curve.

With the same macrospin model, the dependency of $\partial m_2 \chi_{22}^\sigma/\partial H_1$, which determines the sensitivity of the ΔE-effect sensors (Eq. (3)) on the bias field $\mu_0 H_{dc}$ is calculated and plotted in

Figure 1a. In contrast to $\partial \chi_{12}^\sigma/\partial H_1$, $\partial m_2 \chi_{22}^\sigma/\partial H_1$ increases almost quadratically with $\mu_0 H_{dc}$, which results in the typical dependency of $\partial f_r/\partial H$ on $\mu_0 H_{dc}$ for such samples.[13,21] The tilt of the EA causes the shift of the minimum of the $\partial m_2 \chi_{22}^\sigma/\partial H_1$ curve.

Fig 1a highlights the main difference between the ME sensor concepts whose sensitivity is determined by the magnetostrictive strain λ and those whose sensitivity is determined by the magnetization change. Under ideal conditions, in the sensor configuration considered here, the sensitivity is nearly quadratic with $H$ in the first case and linear in the second case. In the following, we show that the ΔE effect contributes significantly to the signal of converse ME sensors, leading to a nontrivial dependency of the sensor signal and bandwidth on the bias field.

*2.1.3. Mixed Modulation. Combination of the Converse ME and ΔE Effects*

To include the contribution of the ΔE effect to the sensor signal, we need to consider the modulation of the resonance frequency $f_r$ of the cantilever induced by $H_{ac}$. The modulation of $f_r$ results in the shift of the frequency dependency of the carrier magnitude $U_c$ and phase $\arg\{u_c\}$ as illustrated in Figure 1c. Consequently, if the excitation frequency $f_{ex} = f_r(H_{ac} = 0)$ is kept constant, the modulation of $f_r$ leads to amplitude (AM) and phase modulation (PM) of $u_c$ (highlighted with a green arrow in Figure 1c). If the cantilever is excited exactly at its resonance frequency, the corresponding modulation of $\sigma(t)$ (Figure 1d) can be expressed as

$$\sigma(t) = (\sigma_0 + S_a H_{ac}(t)^2) \sin(2\pi f_{ex} t + \varphi_\sigma + S_p H_{ac}(t)). \qquad (6)$$

Here, $\sigma_0$ is the amplitude of σ, $\varphi_\sigma$ is a phase lag between $u_{ex}$ and σ for $|H_{ac}| = 0$, $S_a$ is amplitude sensitivity, and $S_p$ is the phase sensitivity expressed as

$$S_p := \frac{\partial \arg\{\sigma\}}{\partial H} = \frac{\partial \arg\{\sigma\}}{\partial f}\bigg|_{f=f_r} \frac{\partial f_r}{\partial H}\bigg|_{H=H_{dc}}, \qquad (7)$$

The amplitude modulation of σ creates harmonic distortion, visible as additional signal peaks at $f_{ex} \pm 2f_{ac}$ in Figure 1d. For the sake of clarity, we restrict our discussion to the impact of the ΔE effect on the main signal peak and assume $S_a = 0$.

Plugging Eq. (7) into Eq. (4) and Eq. (5) gives the final expression for the output signal that combines both the converse ME and the ΔE effects, as illustrated in Figure 1d

$$u_{\text{out}} = A \frac{\partial}{\partial t} \left[ \sigma_0 \sin(2\pi f_{\text{ex}} t + \varphi_\sigma + S_{\text{p}} H_{\text{ac}}(t)) \left( \chi^\sigma + \frac{\partial \chi^\sigma}{\partial H} H_{\text{ac}}(t) \right) \right]. \qquad (8)$$

$S_{\text{p}}$ can be calculated with, e.g., a finite-element model of the magnetoelectrical resonator, as was described in Ref. [22]. However, to avoid these computationally expensive calculations, $S_{\text{p}}$ can be estimated from measurements via

$$S_{\text{p}} \approx \frac{\partial \arg\{u_{\text{c}}\}}{\partial H} = \frac{\partial \arg\{u_{\text{c}}\}}{\partial f}\bigg|_{f=f_{\text{r}}} \frac{\partial f_{\text{r}}}{\partial H}\bigg|_{H=H_{\text{dc}}}, \qquad (9)$$

where $u_{\text{c}}$ is the measured carrier signal of the sensor. The contribution $U_{\text{s, }\Delta\text{E}}$ of the $\Delta$E effect to $U_{\text{s}}$ can be then estimated as

$$U_{\text{s, }\Delta\text{E}} = \mathcal{L}_{f_{\text{ac}}} \left[ U_{\text{c}} \sin\left(2\pi f_{\text{ex}} t + S_{\text{p}} H_{\text{ac}}(t)\right) \right], \qquad (10)$$

where $\mathcal{L}_{f_{\text{ac}}}$ denotes lock-in demodulation of the time-domain signal and selection of its value at $f_{\text{ac}}$.

## 2.2. Influence of the Mixed Modulation on the Signal and Bandwidth of the ME Sensor

In this section, we will demonstrate how the described contribution of the $\Delta$E effect affects the output signal of a converse ME sensor.

### 2.2.1. Signal

The measured sideband amplitude $U_{\text{s}}$ as a function of the magnetic bias field $H_{\text{dc}}$ is shown in **Figure 2**a. A test field $H_{\text{ac}}$ with $\mu_0 |H_{\text{ac}}| = 10$ nT and $f_{\text{ac}} = 10$ Hz was used for the measurements. Figure 2a shows that $U_{\text{s}}$ reaches its minimum of $U_{\text{s}} = 22$ μV at $\mu_0 H_{\text{dc}} = -0.05$ mT and increases with increasing $|\mu_0 H_{\text{dc}}|$ reaching $U_{\text{s}} = 38$ μV at $\mu_0 H_{\text{dc}} = 0.2$ mT. The contribution of the $\Delta$E effect $U_{\text{s, }\Delta\text{E}}$ estimated from the measurements with Eq. (10) and the corresponding change in the resonance frequency $f_{\text{r}}$ is also plotted in Figure 2a. $U_{\text{s, }\Delta\text{E}}$ follows the same trend as $U_{\text{s}}$, showing its minimum of $U_{\text{s, }\Delta\text{E}} = 2$ μV at $\mu_0 H_{\text{dc}} = -0.05$ mT and reaching a maximum of $U_{\text{s, }\Delta\text{E}} = 36$ μV at $\mu_0 H_{\text{dc}} = 0.2$ mT. The resonance frequency $f_{\text{r}}$ shows the opposite trend, reaching the maximum at $\mu_0 H_{\text{dc}} = -0.05$ mT and decreasing with increasing $|\mu_0 H_{\text{dc}}|$. Such dependency of $f_{\text{r}}$ on $\mu_0 H_{\text{dc}}$ is expected from the $\Delta$E effect for the current configuration of the uniaxial magnetic anisotropy and the applied anisotropic stress axis.[13] The increasing change of $f_r$ with $\mu_0 H_{dc}$ leads to higher $U_{s,\,\Delta E}$ (Eq. 7,10), which exceeds 50% of $U_s$ at $\mu_0 H_{dc} > 0.075$ mT.

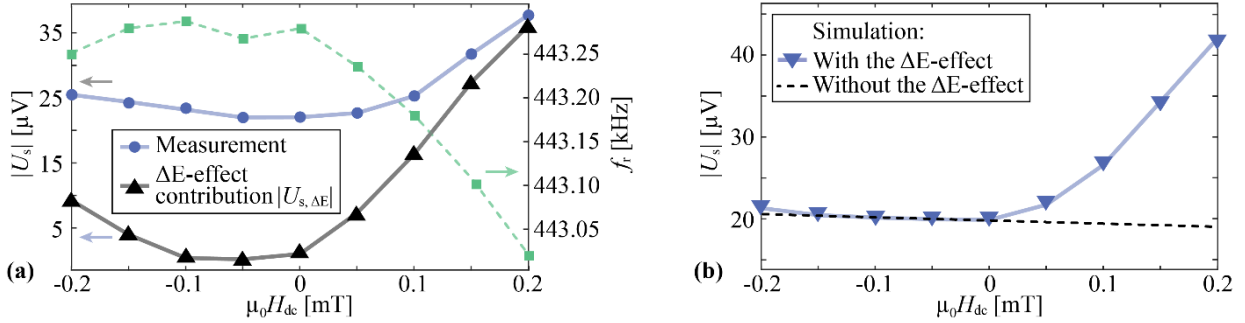

Figure 2. Impact of the ΔE effect on the signal of the converse ME sensor as a function of the bias field $H_{dc}$. (a) Left axis: measured sideband amplitude $|U_s|$ for $|u_{ex}| = 0.1$ V and the contribution $|U_{s,\,\Delta E}|$ of the ΔE effect to $|U_s|$ estimated with Eq. (10). Right axis: resonance frequency $f_r$ of the sensor element. (b) Simulated sideband amplitude $|U_s|$ with and without considering the ΔE effect.

Additionally, we performed a simulation of $u_{out}$ with Eq. (7) using the same sensor signal model as in Ref. [11]. The phase sensitivity $S_p$ is estimated from the measurements using Eq. (9). The sideband amplitude $|U_s|$ was then calculated as $|U_{s,\,\Delta E}| = \mathcal{L}_{f_{ac}}(u_{out})$. The resulting dependencies of $|U_s|$ on $\mu_0 H_{dc}$ for two cases, with and without the ΔE effect, are plotted in Figure 2b. While without the ΔE effect, $|U_s|$ barely changes with $\mu_0 H_{dc}$, with the ΔE effect, $|U_s|$ increases with increasing field magnitude following the increasing $S_p$, as expected from Fig, 1a. This increase of $U_s$ perfectly resembles the trend in the measured data (Figure 2a). Overall, the data indicate that the contribution of the ΔE effect can increase the output signal $|U_s|$ of the sensor system.

*2.2.2. Bandwidth*

According to Eq. (4), the bandwidth of the converse ME sensors should be determined by the frequency dependency of the susceptibility $\chi^\sigma$ i.e., by the magneto-mechanical loss in the magnetostrictive layer.[11] At the same time, the bandwidth of the ΔE-effect sensors is determined by the effective loss of the cantilever, i.e., the mechanical resonator as a whole.[15] Consequently, the bandwidth of the ΔE-effect sensors is narrower in comparison to the converse ME sensors.[6,11] However, if the ΔE effect dominates the signal of the converse ME sensor, the bandwidth of the mechanical resonator should become the dominating factor determining the sensor bandwidth.

To prove this hypothesis, we measured both sensor and the resonator bandwidth over the bias field $\mu_0 H_{dc}$. To estimate the mechanical bandwidth of the resonator $\Delta f_c$, the carrier amplitude $|U_c|$ is measured as a function of the excitation frequency $f$, assuming that $|U_c|$ is directly proportional to the displacement of the cantilever. The result of the measurement performed at $|u_{ex}| = 0.1$ V and $\mu_0 H_{dc} = 0$ mT is plotted in **Figure 3**a. The mechanical bandwidth $\Delta f_c$ of the resonator, estimated as half-width at half-power[23] is $\Delta f_c \approx 100$ Hz.

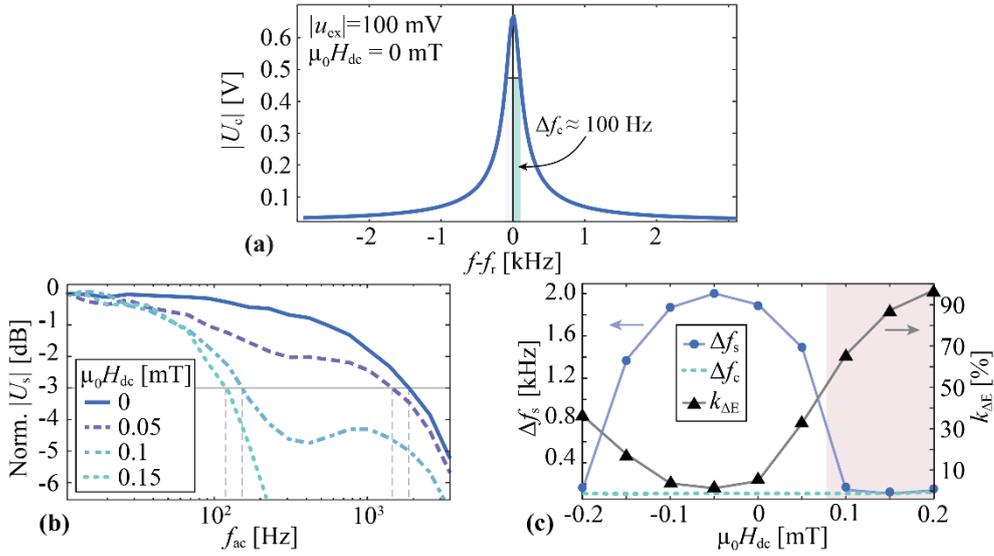

Figure 3. Impact of the ΔE effect on the sensor bandwidth. (a) Carrier heights $|U_c|$ as a function of the excitation frequency $f$. The resonator bandwidth $\Delta f_c$ calculated as half of the half-power bandwidth of $|U_c|$ is highlighted in green. (b) Dependency of the sensor signal $|U_s|$ as a function of the magnetic field frequency $f_{ac}$ recorded at different magnitudes $H_{dc}$ of the magnetic bias field. The values are normalized to $|U_s|(f_{ac} = 10$ Hz$)$ for each $|\mu_0 H_{dc}|$ value individually. (c) Dependency of the sensor bandwidth $\Delta f_s$ and the contribution of the ΔE effect to the sensor signal $k_{\Delta E} = |U_{s, \Delta E}|/|U_s|$ on the bias field $H_{dc}$.

The sensor bandwidth is estimated by sweeping the frequency fac of the applied alternating field $\mu_0 H_{ac}$ with $\mu_0 H_{ac} = 100$ nT and measuring the height $|U_s|$ of the signal peak $f_{ac}$. The resulting dependency $|U_s|(f_{ac})$ normalized to $|U_s|(f_{ac} = 10$ Hz$)$ recorded at $\mu_0 H_{dc} = \{0 \ldots 0.15\}$ mT is plotted in Figure 3b. At $\mu_0 H_{dc} = 0$ mT, the dependency shows the expected low-pass behavior.[25,28] At $\mu_0 H_{dc} = 0.05$ mT and $\mu_0 H_{dc} = 0.1$ mT, the shape of the curve changes, showing two overlapping resonators with different bandwidths.

The resonator with the smaller bandwidth represents the mechanical resonator, and the resonator with the larger bandwidth represents the mechanically driven magnetic resonator. At $\mu_0 H_{dc} = 0.15$ mT, the contribution of the $\Delta E$ effect becomes dominant, and the sensor bandwidth reaches its minimum value equal to the bandwidth of the mechanical resonator.

From the measured dependency $|U_s|(f_{ac}, \mu_0 H_{dc})$ a -3 dB bandwidth $\Delta f_s$ is calculated and plotted in Figure 3b as a function of the bias field $\mu_0 H_{dc}$. For comparison, the contribution of the $\Delta E$ effect to the sensor signal calculated as $k_{\Delta E} = |U_{s,\Delta E}|/|U_s|$ is plotted on the right vertical axis of Figure 3b. Around $\mu_0 H_{dc} = -0.05$ mT, the sensor bandwidth $b$ reaches its maximum of $\Delta f_s = 1990$ Hz. At the same $\mu_0 H_{dc}$, the contribution of the $\Delta E$ effect is the lowest with $k_{\Delta E} \approx$ 1 %. With the increasing contribution of the $\Delta E$ effect, the bandwidth decreases, reaching its minimum level of $\Delta f_s = 120$ Hz at $\mu_0 H_{dc} = 0.15$ mT. For $\mu_0 H_{dc} \geq 0.1$ mT the contribution $k_{\Delta E}$ of the $\Delta E$ effect exceeds 50%, resulting in the narrow sensor bandwidth $\Delta f_s < 150$ Hz (marked red in Figure 3b). The same correlation between increasing $k_{\Delta E}$ and decreasing $\Delta f_s$ is also visible in the negative fields $\mu_0 H_{dc} < 0$. The $\Delta f_s(\mu_0 H_{dc})$ curve is asymmetrical around $\mu_0 H_{dc} = 0$ mT because of the tilt of the magnetic anisotropy axis, as was mentioned before. A clear correlation between the $\Delta E$ effect and the sensor bandwidth is visible, confirming our hypothesis.

## 3. Conclusion

This paper demonstrates that the $\Delta E$ effect plays a crucial role in the performance of magnetoelectric sensors with inductive readout (converse ME sensors). During the operation of such sensors, the $\Delta E$ effect causes amplitude and phase modulation of the stress within the magnetostrictive layer, leading to a mixed modulation of magnetization oscillation. It results in an increase of the signal at low frequencies and a nontrivial frequency dependency of the output signal, leading to a sensor bandwidth that strongly depends on the magnetic bias field.

Notably, in the case of the investigated sensor, the volume fraction of the magnetostrictive material relative to the substrate and other functional layers is about three times smaller compared to reported $\Delta E$-effect sensors.[15,16,22,24] Still, as we have shown, the impact of the $\Delta E$ effect cannot be neglected. This implies that magnetoelastic nonlinearities and other effects connected with the $\Delta E$ effect[22,25] are also relevant for designing converse ME sensors, especially if they are intended to be used in an unshielded environment. Design strategies investigated for ΔE-effect sensors[24,26] could be transferred to converse ME sensors to tune the contribution of the ΔE effect and potentially benefit from this second signal pathway. For suppressing the contribution of the ΔE effect, higher thickness ratios of the substrate to the magnetostrictive layer might be considered. Likewise, to achieve a higher magnetic field tunability of the bandwidth, a larger contribution of the ΔE effect to the signal would be beneficial, which can be achieved by reducing the substrate thickness. Increasing the thickness of the magnetostrictive layer can increase the signal contribution from the converse ME and the ΔE effect.

Additionally, the results show the possibility of tuning the bandwidth and signal output of the sensor system by carefully adjusting operational parameters, which offers an additional dimension to the design of ME sensors. Hence, this work contributes to the fundamental understanding of magnetoelectric magnetic field sensors and marks a significant step towards tailoring magnetoelectric devices to specific applications in the future.

## 4. Methods

*Signal measurement.* The signal measurements were performed with a lock-in amplifier HF2LI from Zurich Instruments and a low-noise operational amplifier LT1363 (from Analog Devices) in non-inverting op-amp configuration. The alternating magnetic field $\mu_0 H_{ac}$ and static magnetic field $\mu_0 H_{dc}$ were created with two coaxial solenoids and two low-noise current sources (6221, Keithley Instruments, Solon, USA). The measurements were performed in a magnetically, electrically, and acoustically shielded environment, including an additional multilayer mu-metal cylinder (Model ZG1, Aaronia).[27]

*Macrospin model.* The macrospin model used for calculating $\chi_{12}^\sigma$ and $\chi_{22}^\sigma$ includes uniaxial magnetic anisotropy, stress-induced anisotropy, and Zeeman energy. Details of the model are provided elsewhere.[11] The saturation magnetostriction constant $\lambda_s$, effective magnetic anisotropy constant $K_u$, and magnetic anisotropy angle $\theta$ relative to $x_2$ are set to $\lambda_s = 30$ ppm,[28] $K_u = 1120$ J/m$^3$, and $\theta = 2.6°$.

**Acknowledgements**

The research was supported by the German Research Foundation (DFG) through the Collaborative Research Centre CRC 1261 "Magnetoelectric Sensors – From Composite Materials to Biomagnetic Diagnostics" and by the Carl Zeiss foundation via the project "Memristive Materials for Neuromorphic Electronics" (MemWerk).

We would like to thank our colleagues Jingxiang Su and Florian Niekiel from Fraunhofer ISITS in Itzeohe for deposition of the piezoelectric AlN and providing the silicon wafers as part of the CRC 1261.